\documentclass[aps,twocolumn,prb,showpacs]{revtex4}

\usepackage{graphicx}

\begin{document}

\title{
The Eigenvalue Analysis of the Density Matrix of 4D Spin Glasses
Supports Replica Symmetry Breaking}

\author{L.~Correale}
\author{E.~Marinari}
\author{V.~Mart\'\i{}n-Mayor} 
\affiliation{Dipartimento di Fisica, 
SMC and UdR1 of INFM and INFN,
Universit\`a di Roma {\em La Sapienza}, 
P. Aldo Moro 2, 00185 Roma, Italy}
\date{\today}

\begin{abstract}
We present a general and powerful numerical method useful to study the
density matrix of spin models.  We apply the method to finite
dimensional spin glasses, and we analyze in detail the four
dimensional Edwards-Anderson model with Gaussian quenched random
couplings.  Our results clearly support the existence of replica
symmetry breaking in the thermodynamical limit.
\end{abstract}

\maketitle

\section{Introduction}

Replica Symmetry Breaking (RSB)~\cite{MPV} was introduced more than
twenty years ago~\cite{PARISI} as a crucial tool to describe the low
temperature phase of spin glasses~\cite{SPINGLASSES}.  One can see
replicas as an extension of Statistical Mechanics that can be very
useful when studying complex systems, such as structural
glasses~\cite{VETRI} or spin glasses~\cite{SPINGLASSES}, where the
ergodicity breaking in the low temperature phase cannot be described
with the help of an infinitesimal external constant magnetic field.

If on one side there is little doubt~\cite{MATHEMATICS} left about the
correctness of the RSB description of the low temperature phase of the
mean field models, on the other side the
controversy~\cite{CONTROVERSIAL,CONTRO-MATH,JSTATPHYS,GROUNDSTATES}
regarding its applicability to finite dimensional systems such as
realistic, physical spin glasses, is alive and in good health.

Unfortunately, we are only starting to guess how to
address the question of the existence of RSB in real spin glasses from
a truly experimental point of view~\cite{FDT}: because of that, and
because of the inherent very high complexity of the relevant analytic
computations, most of the recent progresses are coming from numerical
simulations. 

The output data of numerical simulations are never as
reliable as analytic (and, even better, rigorous)
results. So if on one side the results of numerical simulations of
four dimensional spin glasses~\cite{Zullo,JSTATPHYS} support the RSB
scenario (as indeed happens for the three dimensional
model~\cite{JSTATPHYS}) on the other side one can argue that these
indications could turn out to be fallacious on larger lattices, on
longer time scales, at lower temperatures... (see for example 
\onlinecite{PROBLEMSIMULATIONS} for a typical criticism to typical numerical
simulations).

It is clear that new approaches to this important issue are precious:
Sinova, Canright, Castillo and MacDonald~\cite{SINOVA3} have
recently proposed such a new tool, that can allow a better study of
spin glasses. They have noticed that the spin-spin correlation matrix
$\langle\sigma_i\sigma_j\rangle$ (that we will discuss in better
detail in the next section) shares the main properties of a quantum
mechanical density matrix~\cite{YANG}: it enjoys positivity,
hermiticity and has unit trace (notice that our normalization differs
from theirs, see next section). In the low temperature phase the time
reversal symmetry is broken, and thus one should expect at least one
non vanishing eigenvalue of the density matrix in the thermodynamical
limit, due to the extended character of the eigenvector related with
the symmetry breaking~\cite{YANG}. What is new is the
claim~\cite{SINOVA3} that the presence of RSB is equivalent to
the existence of more than one non vanishing eigenvalues in the
thermodynamic limit. Thus armed the authors of \onlinecite{SINOVA3}
undertook the study of the Edwards-Anderson model with Gaussian
couplings in four dimensions, were they found results that they judged
inconsistent with the detection of RSB on lattices of linear size up
to $6$ (i.e. of volume up to $6^4$). 

The efforts of \onlinecite{SINOVA3} were limited to such small
lattice sizes, because the memory and the numerical effort required in
their approach grows as $L^{2D}$ (in the following $L$ will be the
lattice linear dimension, and $D$ the space dimensionality).  It is
clear that their simulation strategy and data analysis can sometimes
go wrong, as it is evidenced by its failure~\cite{SINOVA4} in the
analysis of the Random Field Ising model. In that case, only turning
to the standard numerical strategy~\cite{JSTATPHYS}, that focuses on
the Parisi order parameter function, $P(q)$, they could
establish~\cite{SINOVA4} the (plausible) absence of RSB in this model.

Here we present a numerical strategy for the study of the density
matrix of spin glass with a cost of the order $L^D$.  We propose a
more convenient data analysis, given the expected behavior of the
density of eigenvalues of the density matrix in the thermodynamic
limit (see next section and reference \onlinecite{CORREALE}). In this
way we have been able to study the Edwards-Anderson model with
Gaussian couplings on lattices of volume up to $8^4$, at the same
temperatures as in \onlinecite{SINOVA3}. We obtain results
that support an RSB scenario~\cite{JSTATPHYS}.  Very interesting
information about the density matrix in a RSB scenario can also be
obtained through mean field calculations~\cite{CORREALE}. Moreover the
numerical approach that we have developed here can be applied to any
spin model.

When completing this manuscript, a note reporting another efficient
approach to the density matrix spectral problem has
appeared~\cite{HUKUSHIMA}. In this work Hukushima and Iba deal with
the four dimensional spin glass model with binary (rather than
Gaussian like in our case) couplings. They have been able to study
lattices of volume $10^4$, reaching the same conclusion that we
present here, i.e.  arguing for the presence of RSB in the infinite
volume limit (they also discuss an interesting method for studying
temperature chaos).

The layout of the rest of this paper is as follows. In
section~\ref{MODELSECTION} we define the model and the associated
density matrix, discussing its basic properties and the numerical
approach of \onlinecite{SINOVA3}. Our own strategy is
presented in subsection~\ref{STRATEGYSECT}, and a working example is
analyzed in subsection~\ref{ISINGSUBSECT}, where the (replica
symmetric) ferromagnetic Ising model in four dimensions is
analyzed. Our numerical simulations of the Edwards-Anderson model in 4
dimensions are described in section~\ref{NUMERICALSECT}. Our results
are presented and discussed in section~\ref{RESULTSSECT}. Finally, we
present our conclusions in section~\ref{CONCLUSIONSSECT}.

\section{The Model and its Density Matrix\label{MODELSECTION}}

We consider the four dimensional Edwards-Anderson spin glass in a
periodic box of side $L$. The $N$ elementary spins can take binary
values, $\sigma_i=\pm 1$, and they are defined on the vertices of a
single hyper-cubic lattice of size $V=L^D$.  We consider a first
neighbor interaction:
\begin{equation}
 H = -\sum_{\langle i,j\rangle}\, \sigma_i J_{i,j}\, \sigma_j\,.
\label{HAMIL}
\end{equation}
The quenched couplings, $J_{i,j}=J_{j,i}$, are drawn from a symmetric
probability distribution function of zero average and variance
$J^2$. It is customary to take $J$ as unit of temperature, and then to
set $J=1$: this is what we do. Two popular choices are the one of a
binary probability distribution $J_{i,j}=\pm 1$ or to take $J$
Gaussian distributed.  Here, we draw the quenched random couplings
from a Gaussian distribution (also in order to allow a direct
comparison with the work of \onlinecite{SINOVA3}).  For all the
relevant observables one first compute the thermal average on a single
realization of the couplings {\em (sample)}, hereafter denoted by
$\langle\ldots \rangle$, and later the average with respect to the
couplings is performed (we denote this disorder average by an
over-line). The model~(\ref{HAMIL}) undergoes a spin glass
transition~\cite{SG4D-GAUSS} at $T_{\mathrm{c}}= 1.80\pm 0.01$.

The average over the couplings $J_{i,j}$ induces a (trivial) gauge
invariance~\cite{GAUGE-SG} in the model. If one chooses a generic
binary value for each lattice site, $\eta_i=\pm 1$, disorder averaged
quantities are invariant under the transformation
\begin{eqnarray}
J_{i,j}&\longrightarrow & \eta_i J_{i,j} \eta_j\,,\\
\sigma_i&\longrightarrow & \eta_i \sigma_i\,.
\label{GAUGECHANGE}
\end{eqnarray}
Now let $\eta_i$ be a random number that takes the probability
$\frac12$ the values $\pm 1$.  If one considers the spin-spin
correlation function, the symmetry~(\ref{GAUGECHANGE}) yields the
disappointing result that
\begin{equation}
\overline{\langle \sigma_i\sigma_j \rangle}= \eta_i\eta_j\overline{\langle \sigma_i\sigma_j \rangle}=\delta_{i,j}\,,
\end{equation}
(that is true since this relation is valid for every value of $\eta$)
explaining why nobody before references \onlinecite{SINOVA3} paid much
attention to this quantity.  Reference \onlinecite{SINOVA3} wisely
suggested to look at the correlation function of a single sample as a
{\em matrix}, $c_{i,j}$. We define here $c_{i,j}$ as
\begin{equation}
c_{i,j} \equiv \frac{1}{L^D}\,\langle \sigma_i \sigma_j\rangle
\label{THEMATRIX}
\end{equation}
(notice the difference in the factor $L^{-D}$ with the definition of
references \onlinecite{SINOVA3,SINOVA4}). The gauge transformation
(\ref{GAUGECHANGE}) acts on the matrix $c_{i,j}$ as an unitary
transformation. Therefore, contrary to the individual elements of
$c_{i,j}$ itself, the spectrum of $c_{i,j}$ does not become trivial
after the disorder average. It is easy to check~\cite{SINOVA3} that
$c_{i,j}$ is symmetric, positive definite, and has trace equal to one,
just like a quantum mechanical density matrix. Thus the corresponding
eigenvalues, $1\ge \lambda_1 \ge \lambda_2 \ge\ldots \lambda_N\ge 0,$
verify
\begin{eqnarray}
1&=&\sum_{k=1}^N\, \lambda_k\,.\label{UNITTRACE}
\end{eqnarray}
Following \onlinecite{YANG} the authors of
\onlinecite{SINOVA3} have argued that in the paramagnetic
phase all the $\lambda_k$ are of order $\frac1N$, and thus vanish in
the thermodynamical limit. On the other hand in the spin glass phase
time reversal symmetry is broken, which implies some non local
ordering pattern for the spins (unfortunately only known by the spins
themselves), and hence at least one eigenvalue, $\lambda_1$ should
remain of order one when $N\to\infty$. They also claimed that presence
of RSB is equivalent to more than one eigenvalue being of order ${\cal
O}(N^0)$ when $N\to\infty$. Furthermore they stated that each non
vanishing eigenvalue corresponds to a pair of pure states: the
correspondence to a {\em pair} of pure states is because of the global
$\sigma\longrightarrow -\sigma$ symmetry of the Hamiltonian
(\ref{HAMIL}) and of the matrix $c_{i,j}$.  Notice that this might be
a clue for the solution of the formidable problem of defining pure
states in a finite volume system~\cite{CONTRO-MATH,JSTATPHYS}.  The
fact that the presence of more than one extensive eigenvalue (of order
${\cal O}(N^0)$) when $N\to\infty$ is equivalent to RSB is true in the
mean field picture, as can be verified in a mean field analytic
computation at the first step of RSB~\cite{CORREALE}.

Combining perturbation theory and droplets ideas
it was also possible to tell~\cite{SINOVA3}  that in a non RSB
scenario the second eigenvalue should not decay slower than
\begin{equation}
\lambda_2 \sim L^{-\theta}\,,
\label{SINOVASCALING}
\end{equation}
where the droplet exponent in four dimensions is~\cite{THETAEXPONENT}
$\theta=0.6$--$0.8$. Actually when the lattice size is larger than the
correlation length (which might not be the case in the achievable
numerical simulations~\cite{PROBLEMSIMULATIONS}), they expect a much
faster decay.

Using the {\em parallel tempering} optimized Monte Carlo
scheme~\cite{TEMPERING}, the authors of \onlinecite{SINOVA3}
calculated the matrix $c_{i,j}$, (a computational task of the order
$L^{2D}$, since the lack of translational invariance prevents the use
of the Fast Fourier transform). They eventually diagonalized the
matrix.  When comparing results for different disorder realizations,
they found very broad distributions of each $\lambda_k$, that they
tried to characterize by their mean and typical value. They found that
the mean and the typical value of the second eigenvalue were
decreasing as a function of lattice size in a double logarithmic plot
for lattices up to $6^4$ (see figure 7 of the second of references
\onlinecite{SINOVA3}). Because of that they argued about the absence
of RSB in the model.

\subsection{An Effective Approach to the Study of the Density Matrix\label{STRATEGYSECT}}

Studying the spin-spin correlation function $c_{i,j}$ by analyzing the
{\em usual} density of states $g_u$
\begin{equation}
g_u(\lambda)=\frac1N \overline{\sum_{k=1}^N \delta(\lambda- \lambda_k)}
\end{equation}
would not work: because of the constraint (\ref{UNITTRACE}) in the
$N\to\infty$ limit $g_u(\lambda)$ is a normalized distribution
function with support in the $[0,1]$ interval with mean value $0$. In
other words, this definition implies that in presence of a generic
finite number of extensive eigenvalues for large volumes
$g_u(\lambda)=\delta(\lambda)$, which does not contain much
information.

In our case we cannot weight all the eigenvalues with the same weight:
to consider a sensible indicator we can decide to use as weight
$\lambda_k$ itself, and to define the modified density of states
of the matrix $c_{i,j}$:
\begin{equation}
g(\lambda)=\overline{\sum_{k=1}^N \lambda_k\, \delta(\lambda-
\lambda_k)}\,.
\end{equation}
It is natural to expect $g(\lambda)$ to converge in the $N\to\infty$
limit to a function containing a continuous part, plus a delta
function at $\lambda=0$ (because a number of order ${\cal O}(N)$ of
eigenvalues will be of order ${\cal O}(N^{-1})$). A calculation at one
step of RSB~\cite{CORREALE} tells us that this is indeed the
case. Moreover in the one step calculation the continuous part do not
show any gap, and it covers all the interval between $\lambda=0$ and
$\lambda=q_{\mathrm{EA}}$, the Edwards-Anderson order parameter (see
also figure 1 of the second of reference
\onlinecite{SINOVA3}). Therefore, from the point of view of checking
Replica Symmetry Breaking, to concentrate on the behavior of
individual eigenvalues does not look the best strategy.  Instead, as
it is customary when analyzing density of states~\cite{ORTHOGONAL},
one can start by considering the moments for a single disorder
realization, $g_J(\lambda)$:
\begin{equation}
\int_0^1\, d\lambda\, \lambda^r g_J(\lambda)= \sum_{k=1}^N \lambda_k^{r+1} =
 \mathrm{Tr}\, c^{r+1}\,.~\label{THEEQUATIONNAIVE}
\end{equation}
Our main
observation is that we can compute the trace of the
$r$-th power of the matrix $c$, using $r$ real replicas (independent
systems, with the same realizations of quenched random 
couplings $J_{i,j}$). 
Let us define the overlap between replicas $a_l$ and $a_j$:
\begin{equation}
q^{a_l,a_j}\equiv\frac{1}{N}\,\sum_{i=1}^N\, 
\sigma_i^{(a_l)} \sigma_i ^{(a_l)}\,.
\label{TRACIAREPLICHE}
\end{equation}
Then it is easy to show that
\begin{equation}
 \mathrm{Tr}\, c^{r} = 
\langle q^{a_1,a_2} q^{a_2,a_3}\ldots q^{a_r,a_1}  \rangle\,.
\end{equation}
Thus, for instance, the (disconnected) spin glass susceptibility is
$\chi_{\mathrm{SG}}= N \ \mathrm{Tr}\, c^{2}$. In this language the
relationship between non vanishing eigenvalues and the phase
transition from the paramagnetic to the spin glass phase is very
direct.

It is now very easy to suggest a numerical strategy of order $L^D$:
make the Monte Carlo simulation in parallel for a discrete number of
replicas, and use them to calculate the appropriate number of moments
of $g_J(\lambda)$. Then use this information to extract the largest
eigenvalues of the matrix $c$.  Unfortunately standard methods for
extracting the probability density from its moments~\cite{ORTHOGONAL}
use orthogonal polynomials. Clearly, given the limited numerical
accuracy that we can expect to obtain for the $\mathrm{Tr}\, c^{r}$,
the use of orthogonality methods is out of the question. We have
instead used a cruder method. We define a cost function
\begin{equation}
{\cal F} (\xi_1,\ldots,\xi_r)= \sum_{l=1}^r\,\left(1\ -\ 
\frac{\sum_{k=1}^r\, \xi_k^l}{\mathrm{Tr}\, c^{l}}\right)^2\,,
\label{COSTFUNCTION}
\end{equation}
and minimize it, using the values of the $\xi_k$ at the minimum as an
approximation to the eigenvalues. This method can be checked on small
lattices, using the direct computation of $c$ and its eigenvalues. It
turns out (see subsection~\ref{ISINGSUBSECT} and section
\ref{RESULTSSECT}) that it is extremely precise for the first
eigenvalue, $\lambda_1$, but that already for the second eigenvalue,
$\lambda_2$ the systematic error is at the $10\%$ level using $12$
replicas. Fortunately we can do better than setting
$\lambda_2\approx\xi_2$.  Let us define a (further) modified density
of states in which we do {\em not} include the first eigenvalue
\begin{equation}
\tilde g(\lambda)=\overline{\sum_{k=2}^N \lambda_k\, \delta(\lambda-
\lambda_k)}\,.
\end{equation}
Its moments are
\begin{equation}
\int_0^1\, d\lambda\, \lambda^r \tilde g(\lambda)=
\overline{\left[\mathrm{Tr}\, c^{r+1} -
\lambda_1^{r+1}\right]}\equiv\overline{\Delta_{r+1}}\,,~\label{THEEQUATION}
\end{equation}
where we have denoted by $\Delta_r$ the subtracted traces.  The right
hand side of equation (\ref{THEEQUATION}) can be accurately calculated
using the cost function, and contains all the information that we
need. 

One could still worry about the bias induced by our use of the cost
function to obtain $\lambda_1$. This can be easily controlled,
because, since the eigenvalues of the matrix decrease fast with $k$ it
turns out that we are always in a situation where we can expect that
$\overline{\Delta_r}$ is clearly and substantially larger than
$\overline{\Delta_{r+1}}$.  On the other hand, if the bias on
$\lambda_1$ is $\delta$, it will affect $\overline{\Delta_r}$ of a
quantity of the order of $(\delta\ r\ \lambda_1^{r-1})$. Therefore, a
bias dominated subtracted trace will be characterized by successive
moments of $\tilde g(\lambda)$ being very similar (see
subsection~\ref{ISINGSUBSECT}).

Let us conclude this subsection by discussing the different scenarios
that could describe the scaling of the subtracted traces, in the
$L\to\infty$ limit.  For a standard replica symmetric model, such as
the usual ferromagnetic Ising model, we expect
$\overline{\Delta_{r+1}} = {\cal O}(L^{-rD})\,$.  In a RSB scenario we
expect that for $L\to\infty$ $\overline{\Delta_{r+1}}$ tends to a
finite value (and that finite volume corrections due to the
eigenvalues that create the $\delta(\lambda)$ in $g(\lambda)$ are of
the form ${\cal O}(L^{-rD})$, while iother finite size corrections due
to critical fluctuations may not decay so fast). Finally, in a droplet
scenario, if one assumes that the subtracted traces are controlled by
$\lambda_2$, then equation (\ref{SINOVASCALING}) implies that
\begin{equation}
 \overline{\Delta_r}={\cal O}(L^{-r\theta})\,,
\label{DROPLETSLAW}
\end{equation}
with $\theta=0.6$--$0.8$ in four dimensions (recall that this is an
upper bound in the decay of $\lambda_2$). The only way out from this
scaling behavior in a droplet picture would be to assume that a number
of the order $L^\xi$ ($\xi>0$) of eigenvalues is of order $L^{-\theta}$: we
are not aware of any argument~\cite{SINOVA3} that would imply the
existence of a divergent number of critical eigenvalues in a droplet
picture.

\subsection{A Simple Example: the Ferromagnetic Ising Model\label{ISINGSUBSECT}}

As a first check we have studied the ferromagnetic Ising model in four
dimensions.  Here the Hamiltonian has the same form than in
(\ref{HAMIL}), but with $J_{i,j}=1$.  We have studied the system at
$T=0.5 T_{\mathrm{c}}$, to prove the deep broken phase with small
correlation length (the critical temperature is here~\cite{ISING4D}
$T_{\mathrm{c}}=6.68025\pm 0.00004)$). We have simulated in parallel
(in this case without parallel tempering, but with an usual heat-bath
updating scheme) twelve replicas of lattices of linear size $L=3,4,6$
and $8$, for $3\times 10^5$ Monte Carlo steps, starting from a fully
ordered state.

In this simple case the density matrix $c_{i,j}$ can be very easily
diagonalized. The correlation function $\langle \sigma_i
\sigma_j\rangle$ depends only on the distance between the two spins,
$\vec x_i - \vec x_j$, and thus the eigenvectors are proportional to
$\mathrm{exp} [ \mathrm{i} \vec k \cdot \vec x_i]$, where the
wave-vectors $\vec k$ verify the usual quantization rules on a
periodic box. It is straightforward to show that the corresponding
eigenvalues are
\begin{equation}
\lambda_{\vec k} = \left\langle\,\left| \sum_{i=1}^N\,
\frac{\mathrm{e}^{\mathrm{i} \vec k \cdot \vec x_i} \sigma_i}
{L^D}  \,\right|^2\,\right\rangle\,,
\end{equation}
and, given the ferromagnetic character of the interaction, the largest
eigenvalue corresponds to $\vec k=0$ (the magnetization, $M$):
\begin{equation}
\lambda_1= \langle M^2 \rangle\,.
\end{equation}

\begin{figure}[htb]
\includegraphics[angle=0,width=\columnwidth]{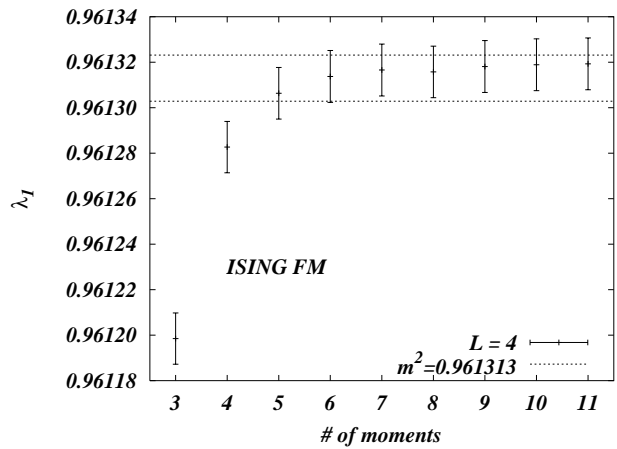}
\caption{Cost function (\ref{COSTFUNCTION}) estimate of the largest
eigenvalue of the density matrix, as a function of the number of
calculated moments (see equation (\ref{THEEQUATION})), for the four
dimensional Ising model at $T=0.5 T_{\mathrm{c}}$, in a $L=4$
lattice. The horizontal lines correspond to $\langle M^2\rangle$ plus
or minus one standard deviation.}
\label{CHECKLAMBDA1-ISING-L4}
\end{figure}

\begin{figure}[htb]
\includegraphics[angle=0,width=\columnwidth]{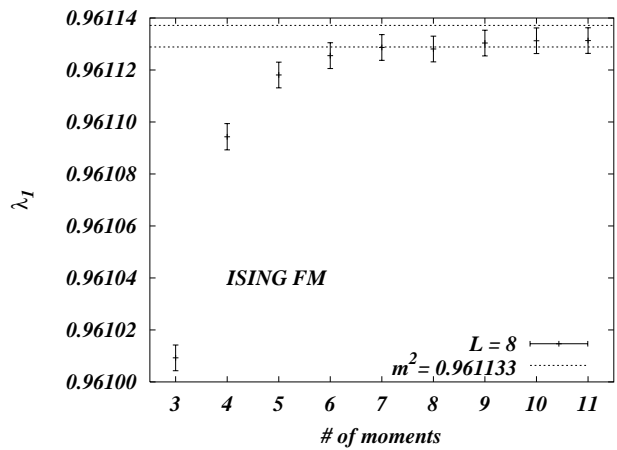}
\caption{
As in figure \ref{CHECKLAMBDA1-ISING-L4} but for a $L=8$ lattice.}
\label{CHECKLAMBDA1-ISING-L8}
\end{figure}

In figures \ref{CHECKLAMBDA1-ISING-L4} and \ref{CHECKLAMBDA1-ISING-L8}
we compare our estimate of $\lambda_1$ for the $L=4$ and $L=8$
lattices, as obtained from the magnetization (the horizontal band is
$\langle M^2 \rangle$ plus/minus an standard deviation), and from the
cost function (\ref{COSTFUNCTION}). As both figures show, 12 replicas
are surely enough to obtain agreement within errors, which in this
case are particularly small.  

\begin{figure}[htb]
\includegraphics[angle=0,width=\columnwidth]{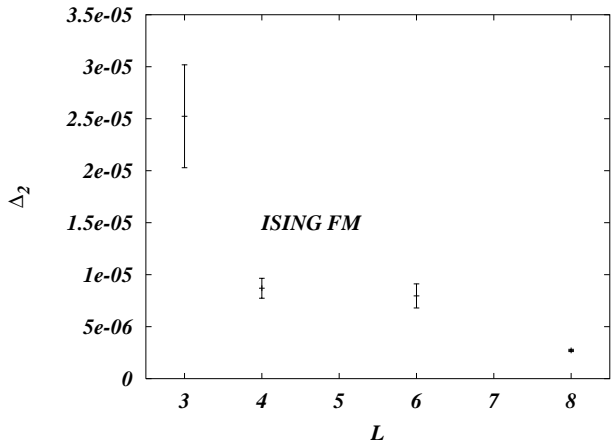}
\caption{The subtracted trace, $\Delta_2$, as a
function of the lattice size, for the four dimensional
ferromagnetic Ising model.}
\label{TRACCIATILDE-2-ISING-L}
\end{figure}

\begin{figure}[htb]
\includegraphics[angle=0,width=\columnwidth]{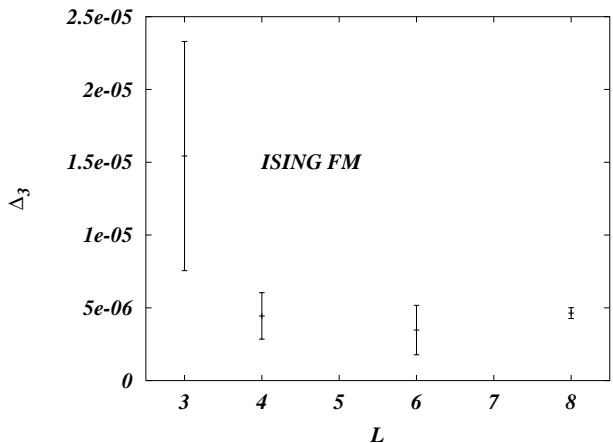}
\caption{
As in figure \ref{TRACCIATILDE-2-ISING-L} but for 
$\Delta_3$.}
\label{TRACCIATILDE-3-ISING-L}
\end{figure}

Having gained confidence in our procedure we can now check evolution
of the subtracted traces with increasing lattice size (figures
\ref{TRACCIATILDE-2-ISING-L} and \ref{TRACCIATILDE-3-ISING-L}). The
two values are very small, decreasing with the lattice size and almost
(but not completely) compatible with zero. One should notice that
$\Delta_3$ and $\Delta_2$ are compatibles within errors for all
lattice sizes (we will see in section~\ref{RESULTSSECT} that in the
spin glass case the situation is very different): in the ferromagnetic
case the real $\Delta_3$ and $\Delta_2$ are so small that they are
completely dominated by the bias discussed in the previous
subsection. One might ask how come that we were able to resolve such
an small bias, given the comparatively large errors reported in
figures \ref{CHECKLAMBDA1-ISING-L4} and \ref{CHECKLAMBDA1-ISING-L8}:
this is due to the strong statistical correlations between$\mathrm{Tr}
(c^r)$ and our estimate for $\lambda_1^r$.

\section{The Monte Carlo Simulation\label{NUMERICALSECT}}

We have studied by numerical simulations the four dimensional
Edwards-Anderson spin glass with quenched random Gaussian couplings
(\ref{HAMIL}).  We have simulated $12$ real replicas in parallel using
a heath bath algorithm and {\em Parallel Tempering} \cite{TEMPERING},
on lattices of volume $3^4$, $4^4$, $6^4$ and $8^4$. The ratio between
full lattice heat bath sweeps and parallel tempering temperature swap
attempt was one to one. For all lattice sizes the largest temperature
was $T_{\mathrm{max}}=2.7$ and the lowest temperature
$T_{\mathrm{min}}=0.8$ (see table~\ref{DETAILS} for details of the
numerical simulation). The probability of accepting a temperature swap
was kept at the $60\%$ level. For each replica we have measured the
permanence histogram at each temperature, and we checked its
flatness. We controlled thermalization by checking that there was no
residual temporal evolution in the Binder cumulant and in
$\mathrm{Tr}\, c^{12}$.

The main scope of the simulation has been to obtain $\mathrm{Tr} c^r$,
for $r=2,\ldots,12$, using equation (\ref{TRACIAREPLICHE}). There is
an awfully large number of equivalent ways of forming the trace
$q^{a_1,a_2} q^{a_2,a_3}\ldots q^{a_r,a_1}$ when one may choose the
replica labels $a_i$ out of twelve possible values.  One needs to find
a compromise between loosing statistics and wasting too much time in a
given disorder realization (the disorder average is the critical
factor controlling statistical error). Our compromise has been the
following: given the special importance of this
observable~\cite{JSTATPHYS} we have calculated the $\frac{12
(12-1)}{2}$ possible overlaps $q^{a_1,a_2}$, and we have computed
$\mathrm{Tr}(c^2)$ using all the $66$ quantities.  For traces of
higher order we have considered only twelve contributions of the form
$q^{i,i+1}q^{i+1,i+2}\ldots q^{i+r,i}$, for $i=1,2,\ldots,12$ (the
sums are understood modulo 12).

\begin{figure}[htb]
\includegraphics[angle=0,width=\columnwidth]{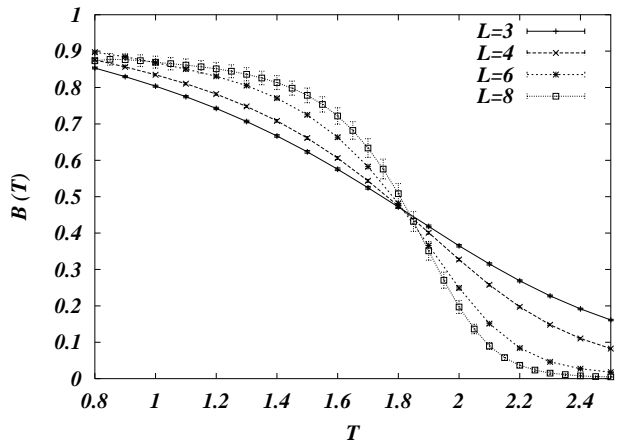}
\caption{The Binder cumulant as a function of temperature, for the
4D Edwards-Anderson model on lattices of linear size
$L=3$, $4$, $6$ and $8$.}
\label{Binder}
\end{figure}

\begin{figure}[htb]
\includegraphics[angle=0,width=\columnwidth]{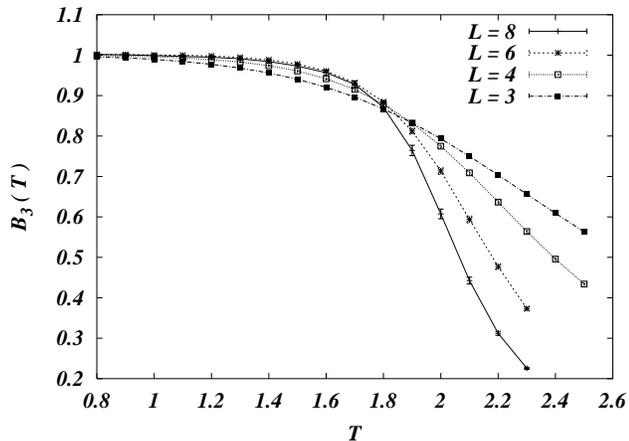}
\caption{As in figure \ref{Binder}, but for the 
$B_3$ cumulant.}
\label{B3}
\end{figure}

In addition to the $\mathrm{Tr}(c^r)$ we have measured the Binder
cumulant (see figure \ref{Binder}) We have also measured a second
adimensional operator
\begin{equation}
B_3 = \frac{\overline{\mathrm{Tr}\, c^3}  }
{\overline{\mathrm{Tr}\, c^2}^{\frac32}}\,,
\label{B3DEF}
\end{equation}
that we show in figure \ref{B3}. 

The theory of finite size scaling~\cite{FSS} predicts that
adimensional quantities close to criticality are functions of
$L^{1/\nu} (T-T_{\mathrm{c}})$, where $\nu$ is the thermal critical
exponent (in $D=4$ one finds~\cite{Zullo} $\nu=1.0\pm0.01)$).  The
crossing points signals the spin glass transition at
$T_{\mathrm{c}}=1.8$ with similar accuracy for both the cumulants that
we have considered. At the lowest temperature that we have reached the
$L=6$ and $L=8$ lattices seem to be far enough from the critical
region.

\begin{table}[b]
\begin{tabular*}{\linewidth}{@{\extracolsep{\fill}}lcccc}
\hline
\hline
$L$ & $N_{\mathrm{samples}}$ & $N_{\mathrm{measures}}$ & 
$N_{\mathrm{thermal}}$ & $N_{\beta}$ \\
\hline &&&&\\
3& 2800 & 50000& 50000& 20\\
4& 2800 & 50000& 50000& 20\\
6& 1208 & 150000& 150000& 40\\
8& 362 & 100000& 200000& 40\\
\hline
\hline
\end{tabular*}
\caption{Relevant parameters of the Monte Carlo simulation. $L$ is the
lattice size. $N_{\mathrm{samples}}$ denotes the number of
realizations of the Gaussian couplings. The number of Monte Carlo
steps (heat bath sweep plus temperature swap attempt) discarded for
thermalization was $N_{\mathrm{thermal}}$.  $N_\beta$ is the number of
temperatures simulated in the parallel tempering. Finally, measures
were taken during $N_{\mathrm{measures}}$ Monte Carlo steps.
\label{DETAILS}}
\end{table}

\section{Numerical Results\label{RESULTSSECT}}

\begin{figure}[htb]
\includegraphics[angle=0,width=\columnwidth]{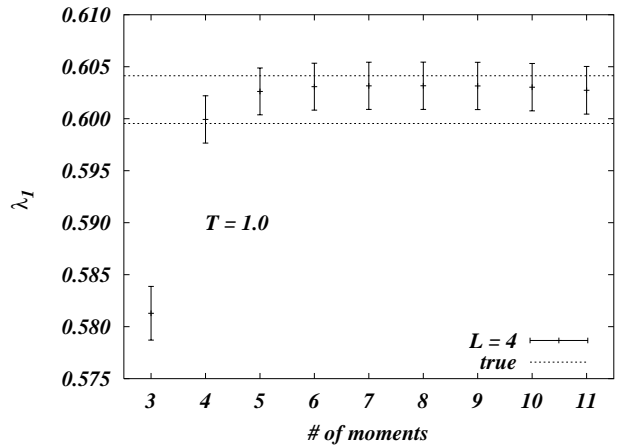}
\caption{Disorder averaged cost function (\ref{COSTFUNCTION}) estimate
of the largest eigenvalue of the density matrix, as a function of the
number of calculated moments (see equation (\ref{THEEQUATION})), for
the four dimensional Edwards-Anderson spin glass at $T=1.0$, on a
$L=4$ lattice. The horizontal lines correspond to a numerical
diagonalization of the matrix $c_{i,j}$ plus or minus a standard
deviation.}
\label{NUMERICALL1}
\end{figure}

\begin{figure}[htb]
\includegraphics[angle=0,width=\columnwidth]{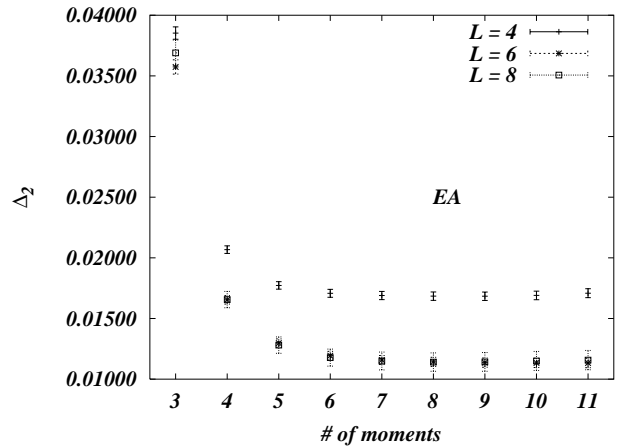}
\caption{Disorder averaged $\Delta_2$ for the four dimensional
Edwards-Anderson spin glass at $T=1.0$ as a function of the number of
computed moments, on different lattice sizes.}
\label{NUMERICALl8}
\end{figure}

\begin{figure}[htb]
\includegraphics[angle=90,width=\columnwidth]{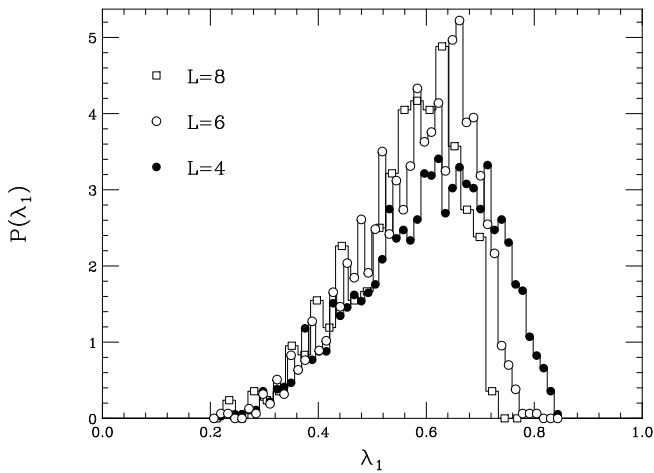}
\caption{Probability distribution of the largest eigenvalue as
calculated in the four dimensional Edwards-Anderson spin glass at
$T=1.0$, for lattices of linear size $L=4$, $6$ and $8$.  The binning
in the $L=8$ lattice was reduced by a factor of two, due to the
smaller number of samples.}
\label{HISTOGRAM}
\end{figure}

To compare our results with the ones of \onlinecite{SINOVA3} we will
specialize here to $T=1.0$. We start by checking on small lattice
sizes (see in figure \ref{NUMERICALL1} the $L=4$ data) the cost
function procedure. In this case the estimate of $\lambda_1$ that one
can obtain by using the cost function can be compared directly with
the result obtained by diagonalization of $c$: we find a fair
agreement. For larger lattices we can only check the convergence of
$\overline{\Delta_r}$ as a function of the number of moments (see
figure \ref{NUMERICALl8}).  Again, the convergence looks fast enough
for our purposes.  We show in figure \ref{HISTOGRAM} the probability
distribution of $\lambda_1$ .  The low eigenvalues tail is basically
lattice size independent.

\begin{figure}[htb]
\includegraphics[angle=0,width=\columnwidth]{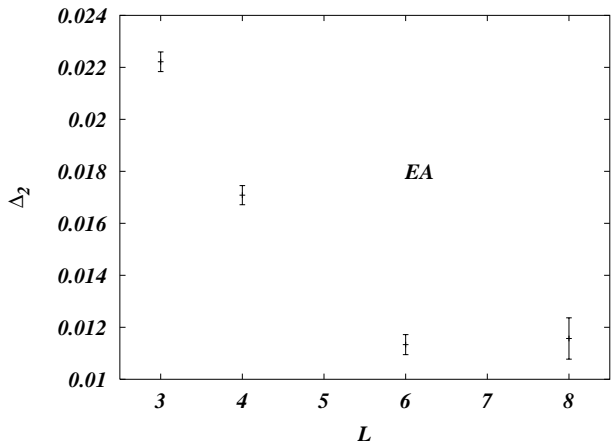}
\caption{Disorder averaged subtracted trace $\overline{\Delta_2}$ for the four
dimensional Edwards-Anderson spin glass 
at temperature $T=1.0$ as a function of the lattice size.}
\label{FINALT2}
\end{figure}

\begin{figure}[htb]
\includegraphics[angle=0,width=\columnwidth]{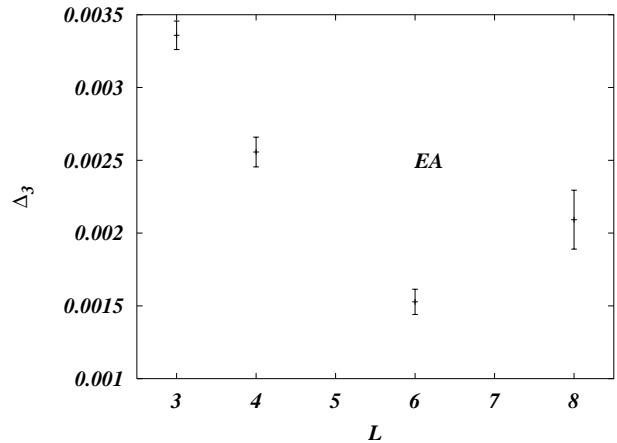}
\caption{As in figure \ref{FINALT2} but for $\overline{\Delta_3}$.}
\label{FINALT3}
\end{figure}

We show
our results for $\overline{\Delta_2}$ and $\overline{\Delta_3}$
in figure \ref{FINALT2} and figure \ref{FINALT3},
respectively. $\overline{\Delta_2}$ is a factor of 10
larger than $\overline{\Delta_3}$: our data are not
bias dominated (see subsections \ref{STRATEGYSECT} and
\ref{ISINGSUBSECT}). The fact that the data point for
$\overline{\Delta_3}$ in the $L=8$ lattice is above the $L=6$ one and
at two standard fluctuations from compatibility may be due either to
a strong fluctuation, or to a first glimpse of bias effects. If one
sticks to the bias hypothesis, the effect on $\overline{\Delta_2}$ can
be (very conservatively) estimated as the difference of the $L=6$ and
$L=8$ data points corresponding to $\overline{\Delta_3}$.  This
difference is well covered by the error in the $L=8$ data point for
$\overline{\Delta_2}$.

\begin{figure}[htb]
\includegraphics[angle=90,width=\columnwidth]{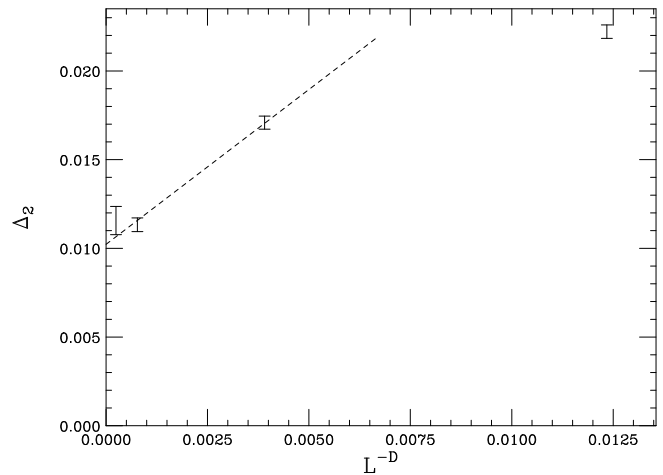}
\caption{Disorder averaged $\overline{\Delta_2}$ as a function of
$L^{-D}$ for the four dimensional Edwards-Anderson spin glass at
$T=1.0$. The dashed line is for a linear best fit, excluding the $L=3$
data.}
\label{RSBFIG}
\end{figure}

\begin{figure}[htb]
\includegraphics[angle=90,width=\columnwidth]{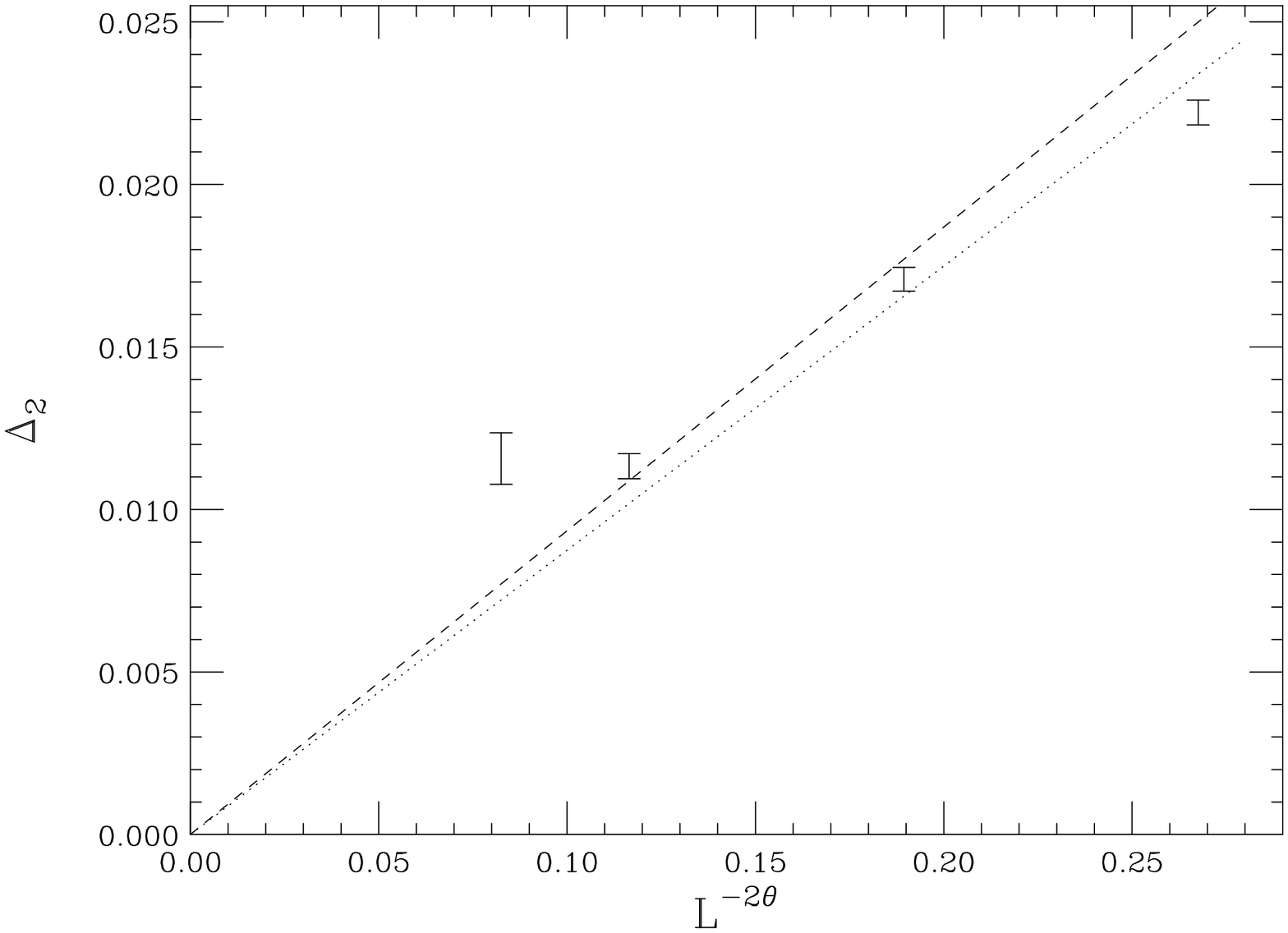}
\caption{Disorder averaged $\overline{\Delta_2}$, as a function of
$L^{-2\theta}$ for the four dimensional Edwards-Anderson spin glass at
$T=1.0$.  The droplet $\theta$ exponent is chosen at its lower bound,
$\theta=0.6$.  The dashed (dotted) line is for a linear best fit,
excluding (including) the $L=3$ data point.}
\label{DROPLETSFIG}
\end{figure}

After the above considerations we can now proceed to the infinite
volume extrapolation. In figure \ref{RSBFIG} we plot the data for
$\overline{\Delta_2}$ as a function of $L^{-D}$.
It is evident that, letting aside the $L=3$ data, a linear fit is
appropriate. The extrapolation to infinite $L$ is definitely
different from zero:  
\begin{eqnarray}
L\ge 3,\ \overline{\Delta_2}=0.0119\pm 0.0003&,&
\chi^2/{\mathrm{dof}}=17.8\,,\\
L\ge 4,\ \overline{\Delta_2}=0.0102\pm 0.0004&,&
\chi^2/{\mathrm{dof}}=1.73\,.
\end{eqnarray}
In figure \ref{DROPLETSFIG} we plot the data as they should scale
according to the droplet model.  A fit to behavior implied by equation
(\ref{DROPLETSLAW}) yields a very high value of $\chi^2/\mathrm{dof}$
either when we include the  $L=3$ data or when we exclude them 
(we use $\theta=0.6$, the
lowest possible value~\cite{THETAEXPONENT}):
\begin{eqnarray}
L\ge 3 &,& \chi^2/{\mathrm{dof}}=17\,,\\
L\ge 4 &,& \chi^2/{\mathrm{dof}}=14\,.
\end{eqnarray}

\section{Conclusions\label{CONCLUSIONSSECT}}

We have proposed and used a new numerical approach to the study of the
density matrix in spin glasses. The original idea of
\onlinecite{SINOVA3}, namely to introduce the density matrix in the
spin glasses context, allows to make interesting
calculations~\cite{CORREALE}, and might even prove useful to the
definition of pure states in finite
volume~\cite{JSTATPHYS,CONTRO-MATH}.

Our method is a further step beyond the useful approach of of
\onlinecite{SINOVA3}. The technology we have developed can be safely
applied to the study of different spin models. The main limitation of
our approach is not related with the use of the density matrix, but
with the extreme difficulty in thermalizing large lattices deep in the
spin glass phase. Should an efficient Monte Carlo algorithm be
discovered, our method would be immediately available, because the
computational burden grows only as $L^D$. Very recently, another
optimized method has been proposed by Hukushima and
Iba~\cite{HUKUSHIMA}. Using their method they were able to study
$10^4$ lattices, using binary rather than Gaussian couplings (which
strongly speeds up the simulation).
 
Using our approach we have been able to show that the density matrix
approach for the four dimensional Edwards Anderson model with Gaussian
couplings in lattices up to $L=8$, and temperatures down to $T=1.0$
($\sim 0.56 T_{\mathrm{c}}$), is fully consistent with an RSB picture,
and that there are serious difficulties with the scaling laws
predicted by the alternative droplet model. In this respect, the
results are in full agreement with the availables
studies~\cite{JSTATPHYS} of the Parisi order parameter, and with the
recent results of \onlinecite{HUKUSHIMA}. A word of caution is in
order: the (postulated) impossibility of getting thermodynamic data in
the reachable lattices sizes~\cite{PROBLEMSIMULATIONS}, affects
equally to the $P(q)$ approach and to the density matrix
approach. However our data for adimensional quantities, such as the
Binder or $B_3$ cumulant, seem very hard to reconcile with the
possibility of a purely finite volume effect.

\section*{Acknowledgments}

We are very grateful to Giorgio Parisi and to Federico Ricci-Tersenghi
for several useful conversations.  Our numerical calculations have been
carried out in the Pentium Clusters RTN3 (Zaragoza), Idra (Roma {\em
La Sapienza}) and Kalix2 (Cagliari). We thank the RTN collaboration
for kindly allowing us to use a large amount of CPU time on their
machine.  VMM acknowledges financial support by E.C. contract
HPMF-CT-2000-00450 and by OCYT (Spain) contract FPA 2001-1813 .

 
\end{document}